\documentclass[a4paper,11pt]{article}
    
\usepackage{jcappub}

\usepackage[T1]{fontenc}
\usepackage[utf8]{inputenc}
\usepackage{graphicx}	
\usepackage{amsmath}	
\usepackage{cleveref}
\usepackage{multirow}
\usepackage{tikz}
\usepackage{comment}
\usepackage{lipsum}
\usepackage{bm}
\usepackage{caption}
\usepackage{float}

\title{Constraining an Einstein-Maxwell-dilaton-axion black hole at the Galactic Center with the orbit of the S2 star}

\author[a]{Rebeca Fernández Fernández,}
\author[b]{Riccardo Della Monica,}
\author[b,c]{Ivan de Martino}

\affiliation[a]{Universidad Internacional de Valencia (VIU), C/Pintor Sorolla 21, 46002, Valencia, Spain }

\affiliation[b]{Universidad de Salamanca, Departamento de Fisica Fundamental, P. de la Merced, 37008 Salamanca, Spain}

\affiliation[c]{Instituto Universitario de Física Fundamental y Matemáticas (IUFFyM), P. de la Merced, 37008 Salamanca, Spain}

\emailAdd{rebeca.ff@hotmail.com}
\emailAdd{rdellamonica@usal.es}
\emailAdd{ivan.demartino@usal.es}

\abstract{
    We derive new constraints on the dilaton parameter appearing in the spherically-symmetric black hole solution of Einstein-Maxwell-dilaton-axion gravity, by studying the geodesic motion of the S2 star in the Galactic Center. Einstein-Maxwell-dilaton-axion black holes represent a compelling alternative to the standard black hole paradigm in General Relativity. This theory emerges from the low energy effective action of the heterotic string theory and has been proven to predict peculiar observational features from the direct imaging of black hole shadows. At a fundamental level, Einstein-Maxwell-dilaton-axion includes additional electromagnetic, dilatonic and axionic fields coupled to the space-time metric. When considering charged non-rotating black hole solutions, the additional fields endow the metric with one extra parameter $b$, called dilaton parameter, that is theoretically bound to $0<b<M$. Using publicly available astrometric data for S2 we derive an upper bound on $b\lesssim 12M$ at 95\% confidence level and we demonstrate that only including the measurement of the relativistic orbital precession for S2 is sufficient to reduce this bound to $b\lesssim 1.4M$ at the same confidence level. Additionally, using a mock data mimicking future observations of S2 with the GRAVITY interferometer, we show that improved astrometric precision can help further narrow down the allowed dilaton parameter range to $b\lesssim0.033M$ after monitoring the S2 orbit for one and a half period.
}

\begin{document}
\maketitle
\flushbottom

\section{Introduction}

The triumph of Einstein's theory of General Relativity (GR) in elucidating the nature of gravity on macroscopic scales has revolutionized our comprehension of space-time and Universe. The astounding predictive power of GR of observational effects beyond the Newtonian description of gravity \cite{Will2014, Berti2015}, such as the perihelion precession of test particles, the deflection of light, the gravitational redshift \cite{Turyshev2008} and the generation and propagation of gravitational waves from the coalescence of compact objects \cite{Bailes2021}, has consolidated its role as a major building block of modern physics.

Probably, among the most striking predictions of GR there is the formation, as a result of gravitational collapse, of black holes (BHs) \cite{Bambi2018}. The existence of these objects, long considered a mere theoretical artifact of GR, is now corroborated by numerous experimental evidences, including the detection of gravitational waves from binary BH mergers \cite{Abbott2016a} and the direct imaging of the so-called `shadow' of the supermassive black holes (SMBHs) M87* at the center of the M87 galaxy \cite{EventHorizonTelescopeCollaboration2019} and Sagittarius A* (Sgr A*) at the Galactic Center (GC) of the Milky Way \cite{EventHorizonTelescopeCollaboration2022a} by the Event Horizon Telescope (EHT). From a theoretical point of view, BHs represent singularities of space-time, where the founding principles of the theory itself break apart \cite{Penrose1965, Hawking1973, Hawking1976}. To avoid such singularities, on very small scales (\emph{i.e.} below the Planck scale of $10^{-35}$ m) GR should emerge from a more complete theory of quantum gravity, which seamlessly integrates gravity with quantum mechanics \cite{Rovelli1996, Schulz2014, Ashtekar2021}. This, along with the shortcomings of GR in explaining observations on galactic and cosmological scales without introducing exotic forms of dark matter and energy \cite{deMartino2020, Li2013}, has led to the formulation of numerous alternative or extended theories of gravity in the endeavour to address in a self-consistent model observational and theoretical flaws of GR \cite{Capozziello2011, Nojiri2017, deLaurentis2023}.

Among the diverse classes of alternatives to the standard BH paradigm, the Einstein-Maxwell-dilaton-axion (EMDA) BHs \cite{Sen1992, GarciaGaltsov1995} have emerged as a subject of significant interest and investigation. These solutions arise as low-energy effective field theories originating from string theory and supergravity \cite{Shapere1991}, offering a compelling framework for unifying the quantum realm and the gravitational interaction. As its name suggests, EMDA includes an interplay between gravity and additional fields (namely Maxwell's electromagnetic field, a dilaton scalar field, and an axion field) which endow this theory with rich dynamics that can profoundly influence the geometry \cite{Sen1992, GarciaGaltsov1995} and thermodynamics \cite{Pradhan2016} of BHs. Since EMDA descends from string theory, and thus incorporates the quantum nature of gravity, peculiar observational signatures that could provide a smoking gun for EMDA BH, and thus an indirect test bench for string theory itself, have been sought after. For this reason, astrophysical implications of EMDA gravity have been studied extensively. For instance, the strong lensing effects leading to the formation of the BH shadow lead to different predictions between EMDA and the standard Kerr BH scenario, that have been thoroughly investigated in \cite{Gyulchev2007, Younsi2016, Mizuno2018, Uniyal2018, Guo2020, Narang2020} resulting in significant differences in the optical appearance of BHs depending on the background metric. In \cite{An2018} the hypothesis that EMDA BHs could serve as accelerators for spinning particles to arbitrarily high energies have been investigated. Furthermore, the continuous spectrum \cite{Banerjee2021a} and jets \cite{Banerjee2021b} from accretion disks around on EMDA BHs have been studied. Recently, the EHT measurement of M87* and Sgr A* have been used to place new constraints on the parameters of EMDA \cite{Sahoo2023}.

In this work we carry out a novel test of EMDA gravity based on the study of the orbits of the S-stars around the four-million-solar-mass SMBH Sgr A* in the GC. In particular, we focus on the motion of the S2 star, the brightest in the cluster, due to its peculiar orbital features \cite{Gillessen2017}. Its short period (of $\sim 16$ years) and high eccentricity (of $\sim 0.88$) result in a very close and fast pericenter passage that enables the detection of special and general relativistic effects on its orbit \cite{GRAVITYCollaboration2018, Do2019, GRAVITYCollaboration2020}. Since the amplitude and impact of these effects strictly depend on the underlying space-time geometry, S2 serves as a direct probe for space-time metric. Here, we aim to leverage this ability to study the current and future ability of astrometric observations of this star to place constraints on the parameters of EMDA gravity.

This article is organized as follows: in Section \ref{sec:EMDA}, we introduce the founding principles of EMDA and of the BH solutions it admits. In Section \ref{sec:methodology}, we describe in details the methodology that we have developed to derive constraints on the parameter space of EMDA. The results of our analysis are reported in Section \ref{sec:results}, and a discussion on how our results insert into the context of astrophysical tests of EMDA is presented in Section \ref{sec:conclusions} along with our conclusions.

\section{Black hole solutions in Einstein-Maxwell-dilaton-axion gravity}
\label{sec:EMDA}
The EMDA theory of gravity arises from the low-energy Lagrangian of the superstring theory \cite{Shapere1991}. In particular, it emerges from the compactification on a six-dimensional torus of ten-dimensional heterotic string theory. What results from this operation is pure ($N = 4$, $d = 4$)-supergravity coupled to ($N = 4$)-super-Yang-Mills theory, in the low-energy limit \cite{Hassan1992, Rogatko2002}. The action $\mathcal{S}$, associated with EMDA gravity contains couplings between the metric $g_{\nu\mu}$, an anti-symmetric electromagnetic tensor field $F_{\mu\nu}$, a $U(1)$ gauge field $A_\mu$, the dilaton field $\chi$ and an axion pseudo-scalar field $\xi$. It can be expressed as \cite{Sen1992, GarciaGaltsov1995}
\begin{align}
\mathcal{S}=\frac{1}{16\pi}\int \sqrt{-g}d^4x \biggl(&R-2\partial_{\nu}\chi\partial^{\nu}\chi-\frac{1}{2}e^{4\chi}\partial_{\nu}\xi\partial^{\nu}\xi+e^{-2\chi}F_{\rho\sigma}F^{\rho\sigma}+\xi F_{\rho\sigma}\tilde{F}^{\rho\sigma} \biggr),
\end{align}
where $g$ is the determinant of the metric, $R$ is the Ricci scalar built from the metric. Such action represents a low-energy effective limit of the heterotic string action up to $\mathcal{O}(\alpha')$  order (being $\alpha'$ the inverse string tension) truncated to contain only those terms involving up to two derivatives. Varying the action with respect to the metric leads to the vacuum field equations
\begin{equation}
G_{\mu\nu}=T_{\mu\nu}(F,\chi,\xi).
\label{eq:field_equations}
\end{equation}
The geometric term $G_{\mu\nu}$ defines the usual Einstein tensor related to the metric $g_{\mu\nu}$, and the right-hand side energy-momentum tensor $T_{\mu\nu}$ contains the effective energy-momentum content of the extra fields in the theory, namely
\begin{align}
T_{\mu\nu}(F,\chi,\xi)=&e^{2\chi} (4F_{\mu\rho}F^{\rho}{\nu}-g{\mu\nu}F^2 )-g_{\mu\nu} ( 2 \partial_\gamma\chi\partial^\gamma\chi+\frac{1}{2}e^{4\chi}\partial_\gamma\xi\partial^\gamma\xi)+\nonumber\\
&+\partial_\mu\chi\partial\nu\chi+e^{4\chi}\partial\mu\xi\partial_\nu\xi.
\end{align}
The field equations \eqref{eq:field_equations} have been demonstrated to admit static BH (black $p$-brane) solutions \cite{Gibbons1987, Garfinkle1991, Horowitz1991, Shapere1991}, that are generally characterized by one or more charges associated with the extra fields. The cancellation of such charges reduces such solutions to the ordinary Schwarzschild BH.
The existence of a rotating stationary and axisymmetric solution for EMDA has been investigated in \cite{Sen1992}. While in the charge-neutral case, the GR Kerr solution is recovered, in the non-null charge case a solution is derived, known as the Kerr-Sen BH, which presents distinctive peculiarities that distinguish it from the charged GR counterpart \cite{Uniyal2018, Mizuno2018, Bernard2016, Ghezelbash2013}. When considering $G=c=1$ and employing Boyer-Lindquist coordinates in a $(-,+,+,+)$ metric signature, this solution reads
\begin{align}
ds^2 = & -\left( 1-\frac{2M\tilde{r}}{\tilde{\Sigma}} \right)dt^2 +\frac{\tilde{\Sigma}}{\Delta}(d\tilde{r}^2+\Delta d{\theta}^2)-\frac{4aM\tilde{r}}{\tilde{\Sigma}}\sin^2\theta dtd\phi+\nonumber\\
&+\sin^2\theta d\phi^2 \left( \tilde{r}(\tilde{r}+r_2)+a^2+\frac{2M\tilde{r}a^2\sin^2\theta}{\tilde{\Sigma}} \right),
\label{eq:metric_rotating}
\end{align}
where
\begin{align}
& \tilde{\Sigma}=\tilde{r}(\tilde{r}+r_2)+a^2\cos^2\theta, \\
& \Delta=\tilde{r}(\tilde{r}+r_2)-2M\tilde{r}+a^2.
\end{align}
Along with the mass $M$ of the central object and the BH spin parameter $a$, the EMDA rotating geometry presents the extra parameters $r_2=\frac{q^2}{M}e^{2\chi_0}$, known as the dilaton parameter. Its value carries information about both the asymptotic value of the dilaton field $\chi_0$ and the electric charge $q$ of the BH, arising from the coupling of the photon with the axion pseudo-scalar. When $q=0$, the Kerr metric is recovered. The geometry in \eqref{eq:metric_rotating} possesses an horizon as long as $r_2$ satisfies the condition
\begin{equation}
0\leq  {r_2} \leq 2M,
\end{equation}
that hence defines the theoretical limits for a BH solution in EMDA. For the purpose of this work (as will be discussed thoroughly in Section \ref{sec:orbital_model}), we can neglect the rotation of the BH and thus consider the non-rotating limit $a\to 0$, where both the spin and the axionic field vanish. A pure dilaton BH solution is hence obtained \cite{Mizuno2018}, given by
\begin{equation}
ds^2=- \left(\frac{\tilde{r}-2\mu}{\tilde{r}+2b}\right)dt^2
+ \left(\frac{\tilde{r}+2b}{\tilde{r}-2\mu} \right)dr^2
+ (\tilde{r}^2+2b\tilde{r})d\Omega^2,
\label{eq:metric}
\end{equation}
Here, $d\Omega^2 = d\theta^2 + \sin^2\theta d\phi^2$ is the solid angle element, and the pseudo-radial coordinate $r$ and mass $M$ are defined as:
\begin{align}
& r^2=\tilde{r}^2 + 2b\tilde{r}, \
& M=\mu+b.
\end{align}
where we have defined a new dilaton parameter $b \equiv r_2/2$, that is theoretically bound to $0\leq b \leq M$ for a BH geometry.

\section{Methodology}
\label{sec:methodology}

In this section, we report our methodology to place constraints on the dilaton parameter ($b$) of an EMDA black hole using astrometric data for stellar orbits at the GC. We provide here details on the dataset used, on the orbital model that we have developed to describe stellar orbits in EMDA, and on the posterior analysis carried out to constrain its parameters.

\subsection{Data}
\label{sec:s2_data}
    
Assiduous near-infrared (NIR) monitoring of the GC during the past 30 years has brought to increasingly precise measurements of the sky-projected trajectories and line-of-sight velocities of the S-stars around Sgr A* \cite{Gillessen2017}. The brightest star in the cluster, S2, with an orbital period of $\sim 16$ yr, is the most easily observed and the least affected by confusion events with other stars in the cluster \cite{Ghez2003}. Astrometric and spectroscopic observations of this star have allowed to determine its Keplerian orbital elements with great accuracy. This allowed to detect, during and after its last pericenter passage in 2018 (at only $120$ AU $\sim 1400 M$ from the central object), departures from a purely Keplerian motion, consistent with relativistic effects tied to the presence of a four-million-solar-mass supermassive compact object \cite{Do2019, GRAVITYCollaboration2018, GRAVITYCollaboration2020}. For all practical purposes, due to the extreme mass ratio with the central object, the S2 star can be regarded as a test particle undergoing geodesic motion in the gravitational field of Sgr~A*. For this reason, the orbit of S2 provides a direct probe of the space-time geometry in the vicinity of the GC that has provided a new avenue to test GR, black hole mimickers, and alternative theories of gravity, on yet unprobed astrophysical scales \cite{DeMartino2021, Borka2021, DellaMonica2022a, DellaMonica2022b, DellaMonica2023b, Cadoni2023}. Motivated by the results in such previous works, here we aim to use the publicly available data for the S2 star to derive constraints on the EMDA BH metric. Although poorly constrained, Sgr A* is likely a rapidly spinning object~\cite{EventHorizonTelescopeCollaboration2022a}. However, following the forecast analysis in~\cite{Grould2017}, we can neglect, at the present sensitivity, the effects on the S2 orbit related to the BH rotation and study the geodesic motion of S2 in the non-rotating metric~\eqref{eq:metric}, to derive constraints on the dilaton parameter $b$ appearing therein. For this purpose, we make use of a dataset comprising astrometric sky-projected positions (right ascension, $\alpha$, and declination, $\delta$) reported in \cite{Gillessen2017} and collected at $N_p = 145$ epochs between $\sim$1992 and $\sim$2017 at different NIR European Southern Observatory (ESO) facilities. These positions are referred to the ``GC radio-to-infrared reference system''~\cite{Plewa2015}, thus requiring the introduction of a possible zero-point offset and drift effect in our orbital model. Information on the motion of S2 over the third dimension, {\textit i.e.} along the observer's line-of-sight (LOS), is provided by $N_v = 44$ measurements of the LOS velocity ($v_{\rm LOS}$) of the S2 star obtained by spectroscopic observations at ESO and reported in~\cite{Gillessen2017}, which cover approximately the same time span. While observations at ESO continued monitoring the S2 star after 2017 with the new and more precise GRAVITY interferometer~\cite{GRAVITYCollaboration2017}, the resulting astrometric dataset is not publicly available and the data at our disposal miss the last S2 pericenter passage in May 2018, where the crucial relativistic effects have been detected~\cite{GRAVITYCollaboration2018, GRAVITYCollaboration2020}. For this reason, we include the information on the measured rate of orbital precession in a different way. In particular, we  add as an additional data point the measurement by the Gravity Collaboration of the relativistic parameter $f_{\rm SP} = 1.10\pm0.19$ which measures the ratio between the observed rate of orbital precession $\Delta \omega$ and its value as predicted by GR:
\begin{equation}
    \Delta\omega_{\rm GR} = \frac{6\pi GM}{ac^2(1-e^2)}.
    \label{eq:precession}
\end{equation}
A value $f_{\rm SP} = 0$ thus corresponds to a Keplerian ellipse that does not suffer from orbital precession and a value $f_{\rm SP} = 1$ corresponds to an amplitude of this effect that is in perfect agreement with a GR BH.

Furthermore, we estimate the impact that future observations with the nominal precision of the GRAVITY interferometer will have on the constraints placed on the EMDA dilaton parameter $b$. In order to do that, we built a mock catalog for S2 (as first presented in \cite{DellaMonica2022a}, to which we refer for further details) with $N_m =141$ sky-projected positions and LOS velocities over one and half orbital period (guaranteeing that we can keep using the non-rotating metric \cite{Grould2017}). These mock data mimic the precision and observational strategy of the GRAVITY interferometer and the SINFONI spectrograph, with nominal uncertainties of $\sigma_A = 10\,\mu$as on the sky-projected positions and $\sigma_V = 10$ km/s for the LOS velocity, respectively \cite{GRAVITYCollaboration2017, Grould2017}.

\subsection{Orbital model for S2 in EMDA}
\label{sec:orbital_model}

Test particles undergoing free-fall motion follow geodesic trajectories of space-time. These four-dimensional curves are solutions of the geodesic equations
\begin{equation}
    \frac{d^2x^\mu}{d\lambda^2} + \Gamma^\mu_{\nu\rho}\frac{dx^\nu}{d\lambda}\frac{dx^\rho}{d\lambda} = 0,
    \label{eq:geodesic}
\end{equation}
where $\lambda$ is an affine parameter (which, for massive test particles, can be chosen as the proper time measured along the geodesic) and the connection coefficients $\Gamma^\mu_{\nu\rho}$ contain derivatives of the space-time metric. This system of second-order ordinary differential equations can be integrated upon setting initial conditions for the four coordinate functions $\{x^\mu\}_{\mu = 0}^3$ and their first derivatives $\{\dot{x}^\mu\}_{\mu = 0}^3$ at a given $\lambda_0$. These 8 degrees of freedom can be reduced by considering that the metric coefficients in both \eqref{eq:metric_rotating} and \eqref{eq:metric} are independent of the coordinate $t$ (so that the value $t_0\equiv t(\lambda_0)$ can be chosen arbitrarily), and that for the case of a time-like geodesic (describing the trajectory of a massive test particle), the four-velocity satisfies the normalization condition $g_{\mu\nu}\dot{x}^\mu\dot{x}^\nu= -1$ which links one of the components of $\{\dot{x}^\mu\}_{\mu = 0}^3$ to the others. For this reason, the space-time trajectory of a test particle is uniquely determined once the three spatial coordinates and the corresponding velocities are assigned. Usually, for the orbits of celestial bodies, these six initial conditions are recast (via the Thiele-Innes transformations) in terms of classic Keplerian orbital elements. Namely, the orbital period $T$ (related to the central mass $M$ via Kepler's third law), the time of passage at pericentre $T_P$, the semi-major axis of the orbit $a$ and its eccentricity $e$ uniquely determine the in-orbital-plane motion of a freely falling object and three angular orbital elements, \emph{i.e.} the orbital inclination $i$, the angle of the line of nodes $\Omega$ and the argument of the pericentre $\omega$ uniquely fix the orbital plane in three-dimensional space. In a general relativistic framework, however, the orbital elements do evolve with time due to post-Newtonian (PN) modifications of the orbital dynamics \cite{Poisson2014}. At first order (1PN), the dominant effect is the precession on the orbital plane of the argument of pericenter $\omega$, given in \eqref{eq:precession} for a Schwarzschild BH. This implies that the trajectory is not a closed ellipse and hence that the orbital elements only identify the ellipse that osculates the trajectory at the initial time.

\begin{figure}
    \centering
    \includegraphics[width = \textwidth]{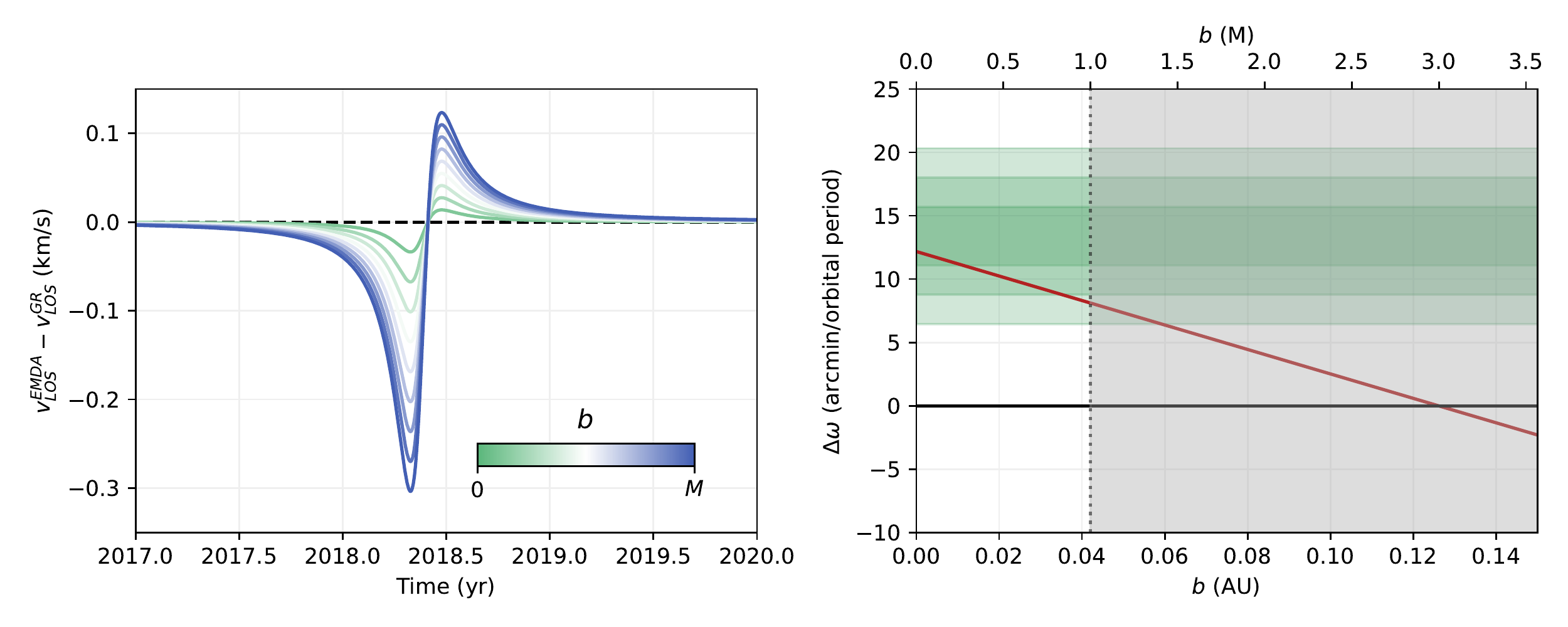}
    \caption{Estimation of the 1PN relativistic effects for the S2 star as a function of the dilaton parameter $b$, computed with our numerical orbital model by fixing all Keplerian elements and only varying $b$. \emph{Left panel:} difference between the special + general relativistic redshift in EMDA and the corresponding effect in GR (obtained by setting $b=0$) around the 2018 pericenter passage for 10 different values of $b$ in the theoretically allowed range $b\in[0,M]$. \emph{Right panel:} Rate of orbital precession for the S2 star in EMDA as a function of the dilaton parameter $b$ (red solid line). The green horizontal shaded bands correspond to the $1\sigma$, $2\sigma$ and $3\sigma$ bounds placed by the analysis in \cite{GRAVITYCollaboration2020}, while the gray shaded region is the theoretically excluded region for a BH solution in EMDA.}
    \label{fig:relativistic_effects}
\end{figure}

Here, we consider as starting point for our integration the last apocenter passage for S2 (in $t_0 \sim 2010.35$), by assigning orbital elements for the initial trajectory at that epoch. We then carry out fully-relativistic numerical integration (both backward and forward in time in order to cover the desired time span) for the geodesic equations \eqref{eq:geodesic} related to the EMDA non-rotating BH metric \eqref{eq:metric} and we build the astrometric observables for the S2 stars, from the integrated orbit. More specifically, three-dimensional positions of S2 around Sgr A* are converted, via Thiele-Innes transformations, into sky-projected angular coordinates ($\alpha,\,\delta$) for an Earth-based observer (which requires fixing the distance $D$ between Earth and the GC). When using actual data, these positions have to be corrected for a possible offset and linear drift over time of the origin of the astrometric reference frame. These effects are described by four additional parameters ($\alpha_0$, $\delta_0$, $v_{\alpha}^0$, $v_{\delta}^0$). The velocity along the LOS is reconstructed by \emph{(i)} considering the LOS-projected kinematic velocity of the star (computed via the same Thiele-Innes transformations); \emph{(ii)} converting the kinematic velocity into a classical longitudinal Doppler effect for the photon frequency emitted by the star; \emph{(iii)} considering additional redshift contributions arising from the 1PN relativistic effects; \emph{(iv)} converting the resulting total frequency shift into the apparent LOS of S2; \emph{(v)} taking into account a possible drift of the LOS velocity, $v_{\rm LOS}^0$ due to the conversion to the local standard of rest (only when considering actual data). In more detail, the 1PN special and general relativistic time dilation effects are related to the high kinematic velocity ($\sim7700$ km/s) and gravitational potential ($\Phi/c^2 = GM/rc^2\sim 3\times 10^{-4}$) of S2 at its pericenter. We estimate this additional redshift contribution directly from the 0-th component of the test particle's four-velocity, $\dot{x}^0 = dt/d\lambda$, from our numerically integrated geodesic, which, by definition, describes the instantaneous rate of coordinate-to-proper time for the test particle and thus takes into account both PN effects at the same time. Finally, we reconstruct the rate of orbital precession $\Delta \omega$ by computing the angle $\Delta \phi$ spanned by the star on the orbital plane between two consecutive radial turning points (\emph{i.e.} points with null radial velocity $\dot{x}^1 = 0$) which by definition is given by $2\pi+\Delta\omega$. In Figure \ref{fig:relativistic_effects}, we report the impact of the two 1PN relativistic effects considered here, redshift and orbital precession respectively, in EMDA for different values of the dilaton parameter $b$ in the range of interest, estimated by means of our numerical orbital model by keeping all other parameters fixed to the ones derived by previous studies \cite{Gillessen2017} and only varying the value of $b$.
As it results, the orbital precession (right panel) can help narrow the bounds for $b$, as evident departure at more than $2\sigma$ (horizontal green stripes) from the purely GR case appears in the considered range for $b$. The gravitational redshift, on the other hand, presents departures with an amplitude of at most $\sim$0.4 km/s from the purely GR, which is well below the current instrumental sensitivity, thus demonstrating the current inability for this relativistic effect to provide any additional constraining power on $b$. Finally, we take into account the classical Rømer delay, due to the different distances of the S2 star at different orbital phases, affecting the apparent position from which the photon started at a given time both in the sky-projected observables and on the LOS velocity.

\subsection{Data analysis}

\begin{table*}
    \centering
	\setlength{\tabcolsep}{13.5pt}
	\renewcommand{\arraystretch}{1.3}
	\caption{The full set of priors used in our analysis.}
	\begin{tabular}{|lcr|}
	\hline
    \textbf{Uniform} & Start & End \\ \hline
    $D$ (kpc) & 6.73 & 9.93 \\
    $T$ (yr) & 15.8 & 16.2 \\
    $T_P$ (yr) & 2018.03 & 2018.63 \\
    $a$ (mas) & 0.1165 & 0.1345 \\
    $e$ & 0.8649 & 0.9029 \\
    $i$ ($^\circ$) & 130.18 & 138.18 \\
    $\Omega$ ($^\circ$) & 220.94 & 232.94 \\
    $\omega$ ($^\circ$) & 59.81 & 71.21 \\
    $b$ ($M$) & 0 & 50 \\
    \hline
    \textbf{Gaussian} & Mean & FWHM \\ \hline
    $\alpha_0$ (mas) & 0.0 & 0.2 \\
    $\delta_0$ (mas) &  0.0 & 0.2 \\
    $v_\alpha^0$ (mas/yr) & 0.0 & 0.1 \\
    $v_\delta^0$ (mas/yr) & 0.0 & 0.1 \\
    $v_{\rm LOS}^0$ (km/s) &  0.0 & 5.0 \\ \hline
	\end{tabular}
	\label{tab:priors}
\end{table*}

The parameter space of our orbital model has been explored with the Bayesian sampler  \texttt{emcee}~\cite{ForemanMackey2013} which carries out a Markov Chain Monte Carlo (MCMC) analysis. Our priors for the Keplerian orbital elements of S2 are uniform probability distributions centered on the best-fit values derived in a previous study on the same dataset in \cite{Gillessen2017} and spanning ranges with amplitudes being 10 times those of the corresponding credible intervals. For the parameters related to the drift and offset of the reference frame, if present in the analysis, we assign Gaussian priors with mean and a full-width half maximum (FWHM) taken from the independent analysis in \cite{Plewa2015}. Finally, for the EMDA parameter, we set uniform priors in a heuristically-fixed interval of $[0,\,50]M$. For the sake of clarity, we report all priors in Table \ref{tab:priors}.
In the first analysis (that hereafter we label as (A)), we fit our orbital model to the astrometric data from \cite{Gillessen2017} alone, meaning that no information about the 2018 pericenter passage of S2 is considered in this analysis. In this case, we sample the posteriors distribution for the full 14-dimensional parameter space, considering all orbital elements and the reference frame parameters, along with the dilaton parameter, as free. For each extracted 14-dimensional sample 
\begin{align}
    \bm{\theta} =& (D,\,T,\, T_P,\, a,\, e,\, i,\, \Omega,\, \omega,\, \alpha_0,\, \delta_0,\, v_\alpha^0,\, v_\delta^0,\, v_{\rm LOS}^0,\, b),
\end{align}
we quantify the agreement between the data and the theoretically predicted positions and velocities with the following log-likelihood function
\begin{align}
    \nonumber\\
	\log \mathcal{L}_{(A)}(\bm{\theta}) =& -\frac{1}{2}\biggl[\sum_i^{N_p}\biggl(\frac{\alpha_i(\bm{\theta})-\alpha^{{obs}}_{i}}{\sigma_{\alpha,i}}\biggr)^2+\sum_i^{N_p}\biggl(\frac{\delta_i(\bm{\theta})-\delta^{{obs}}_{i}}{\sigma_{\delta,i}}\biggr)^2+ \sum_i^{N_v}\biggl(\frac{v_{{\rm LOS}, i}(\bm{\theta})-v_{{\rm LOS}, i}^{obs}}{\sigma_{v_{\rm LOS},i}}\biggr)^2\biggr].
	\label{eq:likelihood_data}
\end{align}
Here, labels $obs$ correspond to the observed quantities from \cite{Gillessen2017} at the $i$-th epoch, while the quantities without superscripts are the predicted ones for a given set of parameters $\bm{\theta}$ at that epoch. The $\sigma$s are the corresponding observational uncertainties.
In the second analysis (that we hereafter label as (B)), we include in our dataset a single data point corresponding to the measurement of the parameter $f_{\rm SP}$ in \cite{GRAVITYCollaboration2020} that encodes the rate of orbital precession for S2. While retaining the same 14-dimensional parameter space, we now consider an additional term in the likelihood, which in this case reads
\begin{align}
	\log \mathcal{L}_{(B)}(\bm{\theta}) =& -\frac{1}{2}\biggl[\sum_i^{N_p}\biggl(\frac{\alpha_i(\bm{\theta})-\alpha^{{obs}}_{i}}{\sqrt{2}\sigma_{\alpha,i}}\biggr)^2+\sum_i^{N_p}\biggl(\frac{\delta_i(\bm{\theta})-\delta^{{obs}}_{i}}{\sqrt{2}\sigma_{\delta,i}}\biggr)^2+ \sum_i^{N_v}\biggl(\frac{v_{{\rm LOS}, i}(\bm{\theta})-v_{{\rm LOS}, i}^{obs}}{\sqrt{2}\sigma_{v_{\rm LOS},i}}\biggr)^2+\nonumber\\
	&\biggl(\frac{f_{\rm SP}(\bm{\theta})-f_{{\rm SP}}^{{obs}}}{\sqrt{2}\sigma_{f_{\rm SP}}}\biggr)^2\biggr].
	\label{eq:likelihood_precession}
\end{align}
Since the analysis in \cite{GRAVITYCollaboration2020} has derived the value of $f_{\rm SP}$ using a PN fit on the same data we use from \cite{Gillessen2017} (plus the crucial observations with the GRAVITY interferometer at the pericenter passage), we considered an additional factor $\sqrt{2}$ in the denominators of \eqref{eq:likelihood_precession}, to avoid double counting data \cite{DeMartino2021}.
In the third and last analysis (that hereafter we label as (C)), we use  a mock catalog mimicking GRAVITY data which allows us to lower the dimensionality of the parameter space. Indeed, due to the different operations of this instrument, no possible offset and drift of the astrometric frame origin with respect to the physical position of Sgr A* is allowed. For this reason, we now explore the following 9-dimensional parameter space
\begin{align}
    \bm{\theta} =& (D,\,T,\, T_P,\, a,\, e,\, i,\, \Omega,\, \omega,\, b)
\end{align}
and define the corresponding log-likelihood as
\begin{align}
	\log \mathcal{L}_{(C)}(\bm{\theta}) =& -\frac{1}{2}\sum_i^{N_m}\biggl[\biggl(\frac{\alpha_i(\bm{\theta})-\alpha^{mock}_{i}}{\sigma_{A}}\biggr)^2+\biggl(\frac{\delta_i(\bm{\theta})-\delta^{mock}_{i}}{\sigma_A}\biggr)^2+\biggl(\frac{v_{{\rm LOS}, i}(\bm{\theta})-v_{{\rm LOS}, i}^{mock}}{\sigma_V}\biggr)^2\biggr].
	\label{eq:likelihood_mock}
\end{align}
Note here that the further assumption has been made that sky-positions and LOS velocities are observed at the same epochs and that the uncertainties for astrometric ($\sigma_A$) and spectroscopic ($\sigma_V$) are constants and always equal to the nominal maximum instrumental precision.

\section{Results}
\label{sec:results}

\begin{figure}
    \centering
    \includegraphics[width = \textwidth]{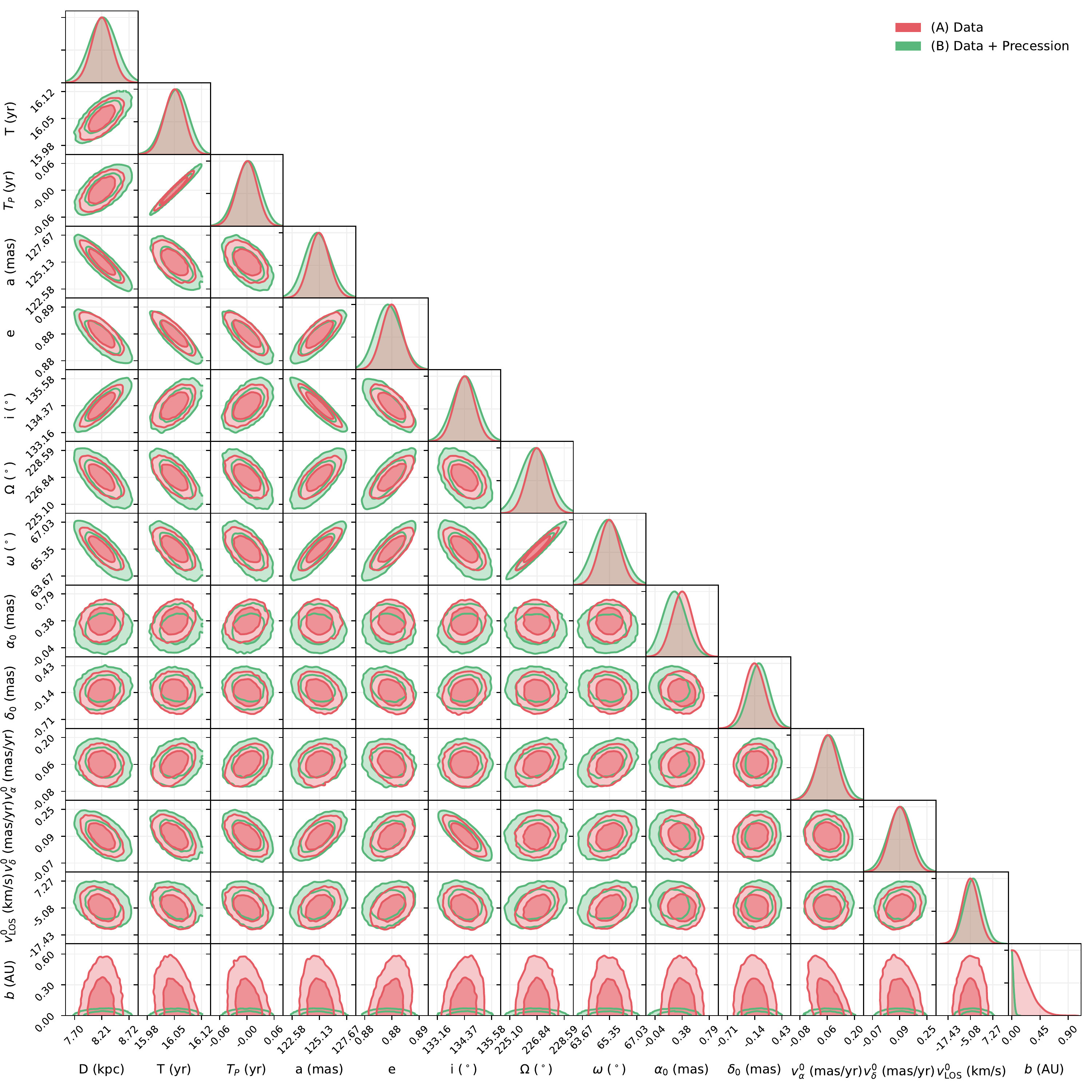}
    \caption{Full posterior distributions for our orbital model for the S2 star in the two analyses with public data. In particular, we depict the 68\% and 95\% confidence regions of the 2D marginalized posterior distribution for each pair of parameters and the histograms, {\it i.e.} 1D marginalized posterior, for the analysis (A) and (B) in red and green, respectively.}
    \label{fig:corner_data}
\end{figure}

\begin{figure}
    \centering
    \includegraphics[width = \textwidth]{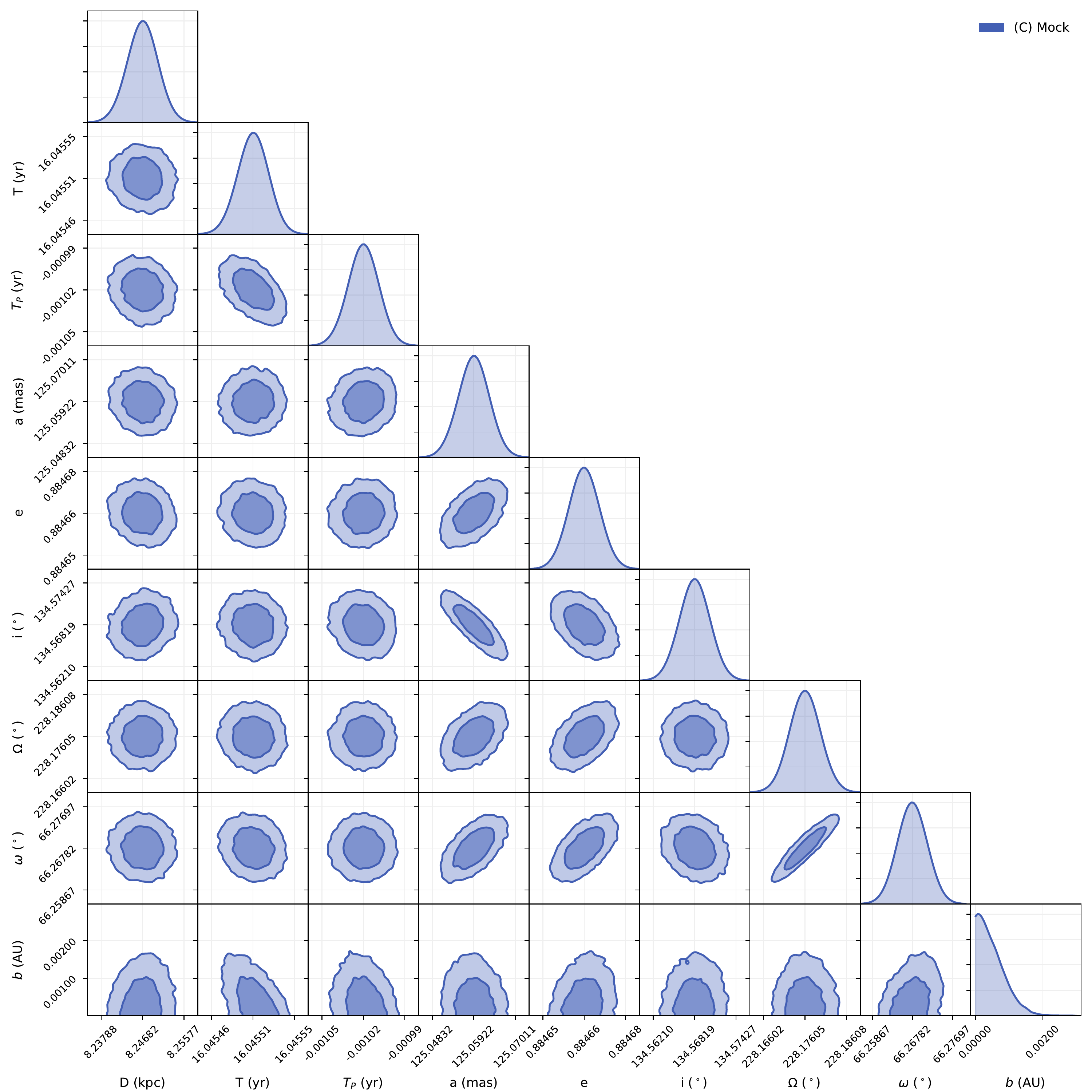}
    \caption{Full posterior distributions for our orbital model for the S2 star in analysis (C).}
    \label{fig:corner_mock}
\end{figure}

\begin{figure}
    \centering
    \includegraphics[width = \textwidth]{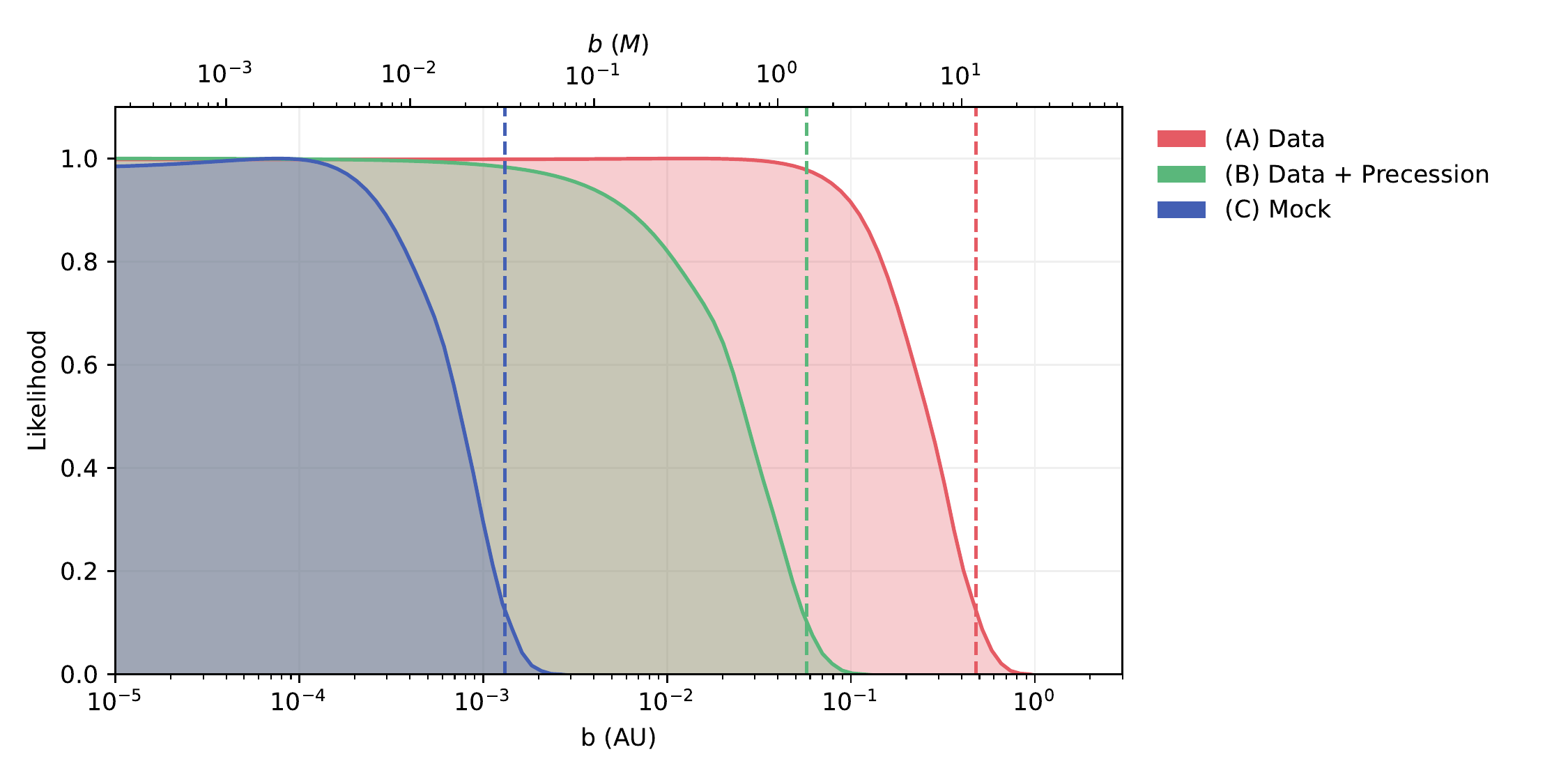}
    \caption{Logarithmic-scale marginalized likelihoods for the EMDA parameter $b$ in the three cases analyzed both in AU (bottom axis) and in units of gravitational radii (top axis). Dashed vertical lines correspond to the respective 95\% confidence upper limits reported in Table \ref{tab:posteriors}.}
    \label{fig:b_posterior}
\end{figure}

\begin{figure}
    \centering
    \includegraphics[width = 0.8\textwidth]{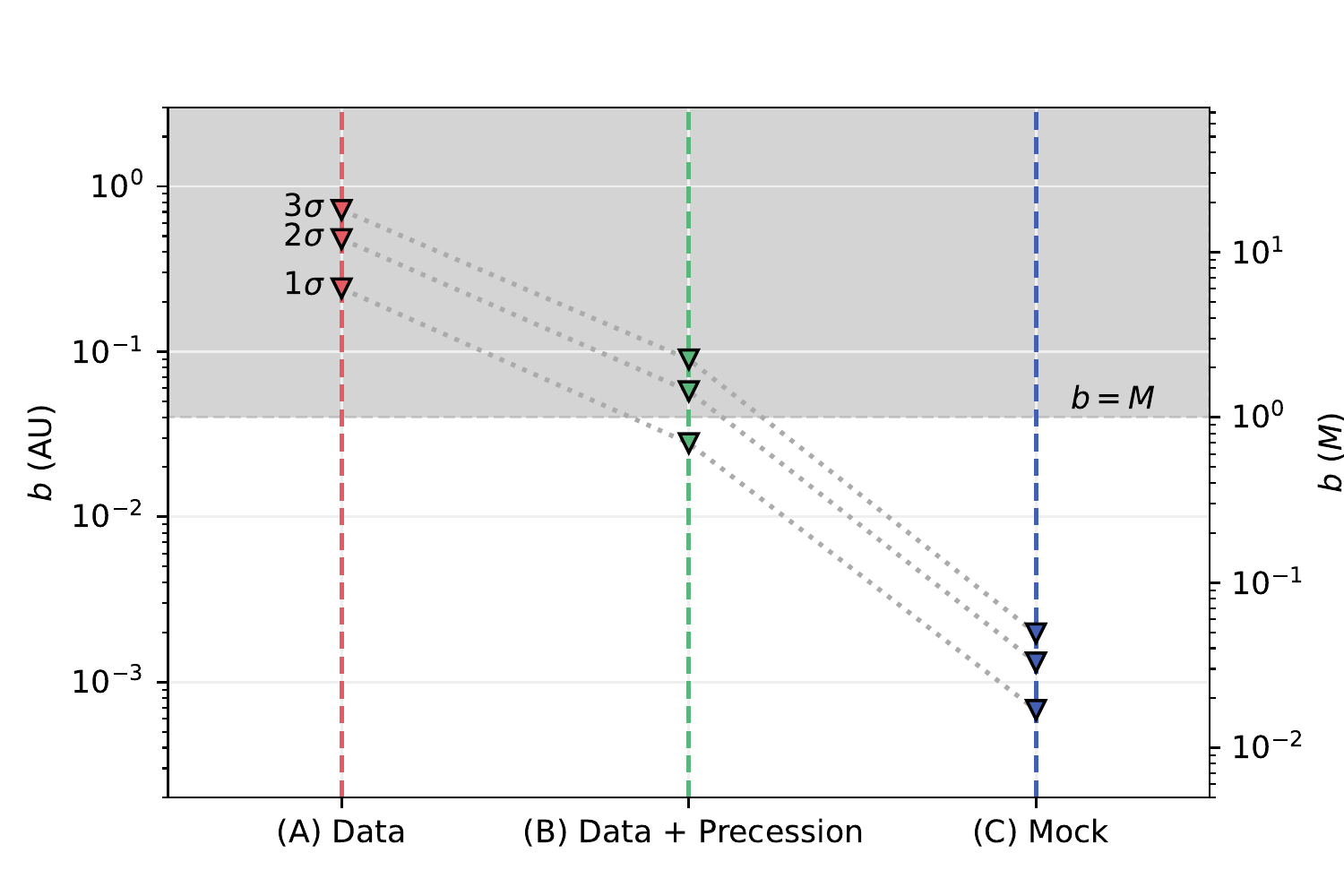}
    \caption{Upper limits at 68\% (1$\sigma$), 95\% (2$\sigma$) and 99.7\% (3$\sigma$) confidence level for the parameter $b$ from our three analyses, sorted in order of increasing constraining power on this parameter. The shaded area corresponds to the theoretically excluded region $b > r_G$.}
    \label{fig:b_upperlimits}
\end{figure}

The results of our three MCMC analyses are reported in Figures \ref{fig:corner_data} and \ref{fig:corner_mock} and summarized in Table \ref{tab:posteriors}. In particular, we show contour plots corresponding to the 68\% and 95\% confidence regions for each pair of parameters, along with marginalized posterior histograms for each single parameter in the orbital model. Figure \ref{fig:corner_data} reports the posteriors from the 14-dimensional analysis (A) (red contours) and (B) (green contours). The two analyses result in  basically overlapping contours for the Keplerian orbital elements of S2, centered on the same values and both compatible with other analyses in literature \cite{Gillessen2017}, except for the marginalized posterior on the parameter $b$ which results much more constrained for the second case, leveraging the constraining power provided by the additional information from the orbital precession. Figure \ref{fig:corner_mock}, on the other hand, reports the posteriors from the 9-dimensional analysis (C). The lower uncertainties on the astrometric observables in our mock catalog result in much narrower confidence regions for all the parameters, including the dilaton parameter $b$, which is much more tightly constrained. All the cases analyzed result in an upper limit for the parameter $b$ and the GR-limit $b = 0$ is always within the $1\sigma$ confidence region. However, due to the different datasets used, the values of the upper limits on $b$ from the different analyses differ by several orders of magnitude. In order to compare such different constraints placed on the dilaton parameter, in Figure \ref{fig:b_posterior} we report the log-scale marginalized likelihoods on $b$ arising from the three analyses with the respective 95\% confidence level upper limits, highlighted by vertical dashed lines. Moreover, Figure \ref{fig:b_upperlimits} shows the upper limits imposed at 68\%, 95\%, and 99.7\% confidence level for the parameter $b$ from our three analyses, sorted in order of increasing constraining power on this parameter and compared to the theoretical bound $b < M$ compatible with a BH space-time. As it results, the analysis performed on public astrometric data alone provides the worst constraints on the parameter $b$ that, at all the confidence levels considered, exceeds the theoretical bound, giving $b \lesssim 12M$ at 95\% confidence. The introduction in our analysis of the information about the orbital precession allows us to bring down the upper limit on $b$ by almost one order of magnitude, resulting in $b \lesssim 1.4M$ at 95\% confidence. While this limit still exceeds the theoretical bound $b<M$, when lowering the confidence level at $68\%$ the same analysis results in an upper limit $b \lesssim 0.8M$, thus limiting the theoretically allowed parameter space for $b$. Finally, the strongest constraints on the EMDA dilaton parameter are derived with our sensitivity analysis using mock data for S2. In this case, we forecast that observations over one and a half orbits for S2 measured at nominal accuracy of 10 $\mu$as would be able to bring down the upper limit on $b$ by almost two orders of magnitude, with an upper bound at 95\% confidence of $b \lesssim 0.033M$, thus narrowing much more tightly the EMDA parameter space. 

\begin{table}
    \centering
    \setlength{\tabcolsep}{9pt}
    \renewcommand{\arraystretch}{1.5}
    \caption{Median values and 68\% confidence interval  for all the orbital parameters of the S2 star as resulting from our three MCMC analyses. The two bottom rows highlight the 95\% confidence level upper limit on our parameter of interest $b$ reported both in AU and in units of gravitational radii ($M$, being $G=c=1$) of the central object.}
    \label{tab:posteriors}
    \begin{tabular}{|l|ccc|}
    \hline
        Parameter (unit) & (A) Data & (B) Data + Precession & (C) Mock \\ \hline
        $D$ (kpc) & $8.21\pm0.17$ & $8.23\pm0.23$ & $8.2468_{-0.0030}^{+0.0029}$ \\
        $T$ (yr) & $16.046\pm0.023$ & $16.050\pm0.028$ & $16.045509\pm0.000015$ \\
        $T_P$-2018.38 (yr) & $-0.003\pm0.020$ & $-0.001_{-0.024}^{+0.023}$ & $-0.0010165_{-0.0000095}^{+0.0000097}$ \\
        $a$ (mas) & $125.13_{-0.85}^{+0.84}$ & $125.0\pm1.1$ & $125.0592\pm0.0036$ \\
        $e$ & $0.8835\pm0.0020$ & $0.8828\pm0.0025$ & $0.8846640\pm0.0000063$ \\
        $i$ ($^\circ$) & $134.37_{-0.41}^{+0.40}$ & $134.40\pm0.50$ & $134.5682\pm0.0020$ \\
        $\Omega$ ($^\circ$) & $226.84\pm0.58$ & $226.75_{-0.84}^{+0.83}$ & $228.1761\pm0.0033$ \\
        $\omega$ ($^\circ$) & $65.35\pm0.56$ & $65.23_{-0.79}^{+0.78}$ & $66.2678_{-0.0030}^{+0.0031}$ \\
        $\alpha_0$ (mas) & $0.38\pm0.14$ & $0.26\pm0.16$ & \\
        $\delta_0$ (mas) & $-0.14\pm0.19$ & $-0.05\pm0.19$ &  \\
        $v_\alpha^0$ (mas/yr) & $0.063\pm0.046$ & $0.072_{-0.052}^{+0.053}$ &  \\
        $v_\delta^0$ (mas/yr) & $0.091\pm0.053$ & $0.094_{-0.063}^{+0.062}$ &  \\
        $v_{\rm LOS}^0$ (km/s) & $-5.1\pm4.1$ & $-3.5\pm4.5$ &  \\ \hline
        $b$ (AU) & $\lesssim 0.48$ & $\lesssim 0.058$ & $\lesssim 0.0013$ \\
        $b$ ($M$) & $\lesssim 12.0$ & $\lesssim 1.4$ & $\lesssim 0.033$ \\
    \hline
    \end{tabular}
\end{table}

\section{Discussion and conclusions}
\label{sec:conclusions}

In EMDA gravity a compelling alternative to the standard general relativistic BH paradigm is formulated. Descending as a low-energy effective limit of string theory, this model has motivated numerous studies in the literature to find peculiar observational signatures that could falsify it against GR \cite{Gyulchev2007, Younsi2016, Mizuno2018, Uniyal2018, Guo2020, Narang2020}. Both charged rotating and non-rotating BH solutions in this theory have been known for a long time \cite{Sen1992, GarciaGaltsov1995} and their properties have been studied extensively \cite{Rogatko2002, Pradhan2016}. The extra fields that enrich the dynamics of this theory, endow the static BH space-time metric with one extra parameter $b$, called the dilaton parameter, whose theoretical bound is $0<b<M$. Past studies in the literature have tried to derive experimental bound on $b$ using astrophysical observations. For instance, in \cite{Banerjee2021a}, using the optical continuum spectrum of the quasars, a preferred value of $r_2 \sim 0.2$ (corresponding to $b\sim 0.1$) is found and the range $r_2 \gtrsim 1.6$ ($b\sim 0.8$) is shown to be disfavoured. In \cite{Banerjee2021b}, on the other hand, the jet power and radiative efficiency of microquasars are used to demonstrate that the region of the parameter space $r_2 \gtrsim 1$ ($b \gtrsim 0.5$) is generally disfavored, with $r_2 = b = 0$, \emph{i.e.} standard GR Kerr, always favored over the EMDA Kerr-Sen black hole. Finally, a recent study in \cite{Sahoo2023} has derived an observational bound on the dilaton parameter using the measured shadow diameter for M87* and Sgr A*. While the shadow of M87* exhibits a preference towards the GR scenario, the shadow of Sgr A* surprisingly shows a preference towards the Kerr-Sen scenario with an allowed interval $0.1 \lesssim r_2 \lesssim 0.4$ ($0.05 \lesssim b \lesssim 0.2$), owing to the smaller shadow cast by BH with larger dilaton parameter.

In this work, we have studied the astrometric observational signatures of EMDA on the trajectory of the S2 star in the GC of the MW. Due to its orbital properties, the orbit of S2 stands out as a direct probe of the space-time geometry in the vicinity of a SMBH, on which the detection of relativistic effects \cite{Do2019, GRAVITYCollaboration2018, GRAVITYCollaboration2020} not only allows for a fresh take on the classical tests of GR \cite{Angelil2011, Qi2021, deLaurentis2023} but also offers a new probe to falsify the standard BH paradigm in GR against possible alternatives \cite{DeMartino2021, Borka2021, DellaMonica2022a, DellaMonica2022b, DellaMonica2023b, Cadoni2023}. By developing a fully relativistic orbital model for S2 in the static BH metric in EMDA, we have derived new constraints on the dilaton parameter $b$. To do that, we have carried out three separate MCMC analyses showing that the credible interval for this parameter can be greatly reduced once one either takes into account the measurement of the rate of orbital precession for S2 \cite{GRAVITYCollaboration2020} or a much improved astrometric accuracy that can be reached in the future with the GRAVITY interferometer. In particular, the first analysis, (A), was performed considering public astrometric data in \cite{Gillessen2017} alone, from which the upper limit $b \lesssim 12M$ has been derived at 95\% confidence. Introducing in analysis (B) one single data point corresponding to the rate of orbital precession has brought down such upper limit on $b$ by almost one order of magnitude: $b \lesssim 1.4M$ at 95\% confidence, demonstrating the great constraining power associated with this relativistic effect. Finally, in the sensitivity analysis (C), using mock data developed in \cite{DellaMonica2022a} mimicking the nominal accuracy and observational strategy of the GRAVITY interferometer, we forecast a tight constraint on $b$ of order $b \lesssim 0.033M$  at 95\% confidence, after one and a half period of observation. All the analyses always recover the GR limit $b = 0$ within the $1\sigma$ confidence interval.

This study has further demonstrated the importance of the observations of stellar orbits at the GC to serve as probes for space-time geometry. These can provide a crucial tool in the future, when much higher precision will be achieved in the single orbit tracking, to leverage the constraining power of relativistic effects in the endeavor of narrowing down the allowed range for departures from the standard general relativistic description of gravity.

\acknowledgments
RDM acknowledges support from Consejeria de Educación de la Junta de Castilla y León. IDM acknowledges support from Grant IJCI2018-036198-I  funded by MCIN/AEI/10.13039/ 501100011033 and, as appropriate, by “ESF Investing in your future” or by “European Union NextGenerationEU/PRTR”. IDM and RDM also acknowledge support from the  grant PID 2021-122938NB-I00 funded by MCIN/AEI/10.13039/ 501100011033 and by “ERDF A way of making Europe”. 

\bibliographystyle{JHEP}
\bibliography{biblio}

\end{document}